[TITLE]

# Transport properties of two laterally coupled vertical quantum dots in series with tunable inter-dot coupling

[AUTHORS & AFFILIATIONS]

T. Hatano[a)], S. Amaha, T. Kubo, S. Teraoka

*JST, ICORP, Quantum Spin Information Project, 3-1, Morinosato-Wakamiya, Atsugi-shi, Kanagawa 243-0198, Japan*

Y. Tokura

*JST, ICORP, Quantum Spin Information Project, 3-1, Morinosato-Wakamiya, Atsugi-shi, Kanagawa 243-0198, Japan* and *NTT Basic Research Laboratories, NTT Corporation, 3-1, Morinosato-Wakamiya, Atsugi-shi, Kanagawa 243-0198, Japan*

J. A. Gupta, D. G. Austing

*Instutute for Microstructural Sciences M50, National Research Council of Canada, Montreal Road, Ottawa, Ontario, K1A 0R6, Canada*

S. Tarucha

*JST, ICORP, Quantum Spin Information Project, Morinosato-Wakamiya, Atsugi-shi, Kanagawa 243-0198, Japan* and *Department of Applied Physics, School of Engineering, University of Tokyo, 7-3-1, Hongo, Bunkyo-ku, Tokyo 113-8656, Japan*

**[ABSTRACT]**

We describe the electronic properties of a double dot for which the lateral coupling between the two vertical dots can be controlled in-situ with a center gate voltage ($V_c$) and the current flows through the two dots in series. When $V_c$ is large and positive, the two dots merge. As $V_c$ is made less positive, two dots are formed whose coupling is reduced. We measure charging diagrams for positive and negative source-drain voltages in the weak coupling regime and observe current rectification due to the Pauli spin blockade when the hyperfine interaction between the electrons and the nuclei is suppressed.

**[KEYWORDS]** vertical quantum dots, charging diagram, Pauli spin blockade, hyperfine interaction

**[TEXT]**

Structures composed of multiple quantum dots (QDs) are expected to have functionality that can not be realized with conventional semiconductor structures. QD structures are being intensively investigated not only for fundamental physics but also for potential quantum information applications [1]. For the latter, structures composed of two coupled QDs can be regarded as the most basis unit (see Ref. [2] and references therein).

Double quantum dot (DQD) structures for transport measurements are commonly constructed from either lateral QDs or vertical QDs [2]. For lateral QDs (coupled laterally), it is non-trivial to tune the voltages on several gates to form the DQD structure in the few-electron regime, and quantum point contact (QPC) detection is often necessary to determine unambiguously the electron number on each QD [3]. On the other hand, for vertical QDs (coupled vertically), although it is straightforward to determine the electron number on each QD without QPC detection, the coupling between the two QDs in the original DQD structures principally depends on the height and width of the hetero-structure barrier between the two QDs, which are fixed during epitaxial growth, so it is not possible to tune the inter-dot coupling strength [4]. Recently, however, DQD structures with laterally coupled vertical QDs have been fabricated which offer gate tunable inter-dot coupling. These structures can be categorized into two types depending on whether the two QDs are arranged in parallel [5] or in series [6] between the source and drain contacts. Concerning the latter type, highly attractive for access to current rectification due to Pauli spin blockade (P-SB) [7], although gate tunable inter-dot coupling capability was incorporated, clear gate tunable DQD behavior and P-SB were not actually observed [6].

In this Letter, we report the electronic properties of a DQD structure with no QPC and four gates only for which the lateral inter-dot coupling between the two vertical QDs arranged in series can be easily tuned with a gate voltage from the strong to the weak coupling regime. In the strong (weak) coupling limit, the measured charging diagram exhibits single-dot-like (weakly-coupled-double-dot-like) characteristics. In the weak coupling regime current suppression in one bias direction due to P-SB is clearly seen when the hyperfine interaction [8,9] between electron spins and nuclear spins is turned off in the presence of a small in-plane magnetic field.

The DQD structure is fabricated from a hetero-structure which consists of a heavily doped GaAs layer on a semi-insulating GaAs substrate, a thick undoped spacer layer, an 80nm $Al_{0.3}Ga_{0.7}As$ lower barrier, a 10nm GaAs well, a 7.5nm $Al_{0.3}Ga_{0.7}As$ upper barrier and a doped $GaAs/Al_{0.05}Ga_{0.95}As$ upper contact region. A scanning electron microscope image of a structure similar to that measured is shown in Fig. 1(a). The two distinct dot mesas (each of size ~400nm) are clear. A schematic cross section (A-B) of the two coupled vertical QDs is given in Fig. 1(b).

Although the hetero-structure is similar to that used to fabricate an earlier DQD structure with two vertical QD arranged in parallel [5], there are two important differences in the DQD structure described in this work. Firstly, the two mesas are physically separate, with the source and drain contacts formed as shown in Fig. 1(b), and so although the inter-dot coupling is still in the lateral direction, the current flows in series through the two QDs rather than in parallel. Secondly, the lower barrier is very thick and so the current can not flow towards the substrate. Accordingly, the current flows in from drain and out to source, both on the top side of the structure. The heavily doped GaAs layer located below the thick lower barrier acts as a buried back gate, and a voltage applied to it allows the carrier density in the well (QD) region to be altered. The structure additionally has three Schottky gates on the surface. Two are side gates, one for each of the mesas, which primarily tune the number of electrons in the two QDs. The main function of the remaining gate, the center gate running between the two mesas, is to control the coupling between the two QDs. All measurements are carried out in a dilution refrigerator.

Figure 2(a) shows the current through the DQD structure as a function of the left and the right side gate voltages, $V_{sL}$ and $V_{sR}$, at fixed center gate voltage $V_c$=0.4V and back gate voltage $V_b$=2.0V, and for constant source-drain voltage $V_{sd}$=100μV. This charging diagram reveals lines, each of which identifies a Coulomb oscillation peak, that run almost in a straight line diagonally from upper-left to lower-right and are parallel to each other. Such characteristics are expected for a single QD and so indicate that the two intended QDs have in fact merged into one large QD [10,11]. For the combination of gate voltages applied, we are unable to empty the large merged QD. From the separation between the Coulomb oscillation peaks and Coulomb diamond measurements (not shown) the intra-dot charging energy in the lower-left region of Fig. 2(a) is estimated to be approximately 1.3meV.

Applying a progressively less positive $V_c$, we expect to split the large merged QD into two QDs initially with relatively strong inter-dot coupling. The corresponding charging diagram for $V_c$=0.35V is shown in Fig. 2(b), and it now takes on the appearance of the familiar honeycomb pattern expected for two coupled QDs [10,11], i.e., there are two sets of approximately parallel lines with different slopes, and lines of different slope anti-cross when they approach each other. The separation between consecutive Coulomb oscillation peaks originating from the charging of the same QD is approximately double the separation in Fig. 2(a) because the large merged QD has split into two QDs of approximately half the size. In general, for two QDs in series, we expect that the current flow is only substantial in the vicinity of the so-called triple points when the energy levels in the left dot and the right dot are aligned. However, for the combination of gate voltages applied, away from the triple-points, the Coulomb oscillation peaks which occur only when the numbers of electrons on the left dot or the right dot change are still observable due to finite co-tunneling. The anti-crossing behavior at the triple points is clear because the coupling energy between the two QDs is comparatively large. The inter-dot coupling energy, including the tunnel and electrostatic coupling energies, in the upper-left region of Fig. 2(b) is estimated to be 300~600μeV [12].

To weaken the coupling between the two QDs, $V_c$ is made less positive again. The corresponding charging diagram for $V_c$=0.3V is shown in Fig. 2(c). Compared to Fig. 2(b), the co-tunneling away from the triple points has become weaker, and towards the lower-left region of Fig.2(c), co-tunneling is essentially absent (with only some "spots" of current remaining at the triple points). Furthermore, the anti-crossing behavior at the triple points is generally less pronounced than in Fig. 2(b), indicating that the coupling between the two QDs has weakened [5,11]. This is particularly evident in the upper-left region of Fig. 2(c). The inter-dot coupling energy in this region is estimated to be ≲200μeV [12].

For DQD structures with the two QDs arranged in series P-SB can occur in the weak coupling regime leading to current rectification [7,9]. This property is highly useful not only for quantum information applications [2], but as a means to examine the hyperfine interaction between electron spins and nuclear spins [2,8,9]. Accordingly, we now look for evidence of P-SB in the transport characteristics of our DQD structure in the weak coupling regime.

Figure 3(a) shows part of a charging diagram [for weaker coupling than that relevant to Fig. 2(c)] in the region where the absolute number of electrons on the left and the right QDs ($N_L$, $N_R$) can be (1,2), (2,2), (1,3), or (2,3) for positive $V_{sd}$=500µV. Note ($N_L$, $N_R$) are determined by measuring the charging diagram over a wide range of $V_{sL}$ and $V_{sR}$, and Coulomb diamonds (not shown). A triangular region within which the current is finite (>1 pA) is clear. Actually, the two expected triangles [one each for the electron and hole cycles involving the transition (2,2)→(1,3)] are so strongly overlapped at $V_{sd}$=500µV that they have effectively merged into one [2,9]. This is because the inter-dot coupling is no more than ~120µeV, and is thus much smaller than $V_{sd}$. Focusing on the same part of the charging diagram as in Fig. 3(a), and for the same values of $V_c$ and $V_b$, Fig. 3(b) shows what is measured for $V_{sd}$=-500µV. The current in the triangular region [again two merged triangles, one each for the electron and hole cycles involving the transition (2,2)←(1,3)] is somewhat different, but not dramatically so, compared to the triangle in Fig. 3(a).

The fact that the triangles in Fig. 3(a) and Fig. 3(b) are not more dissimilar at first sight may seem surprising. To see this, we first note that since all four of the relevant charge states have two "inert" electrons with anti-parallel spins trapped in the lowest single-particle energy level in the right QD, we can essentially neglect this filled "core level" [5,13]. Thus, the four charge states can be equivalently regarded as (1,0), (2,0), (1,1), and (2,1), and so the electron and hole cycles leading to the current in the triangular regions in Fig. 3(a) and Fig. 3(b) respectively effectively include the transitions (2,0)→(1,1) and (2,0)←(1,1) [14]. For the latter, though, we would expect P-SB to lead to strong current suppression because the two dot system can become trapped in the (1,1) triplet state [7,9], and we would expect this too for the (1,3) triplet state in reverse bias [Fig. 3(b)]. However, this expectation has neglected the hyperfine interaction between the electron spins and the nuclear spins in the host material of the two dots which allows triplet and singlet states to mix so lifting P-SB [7,9]. In the weak coupling regime appropriate for our DQD structure, we expect hyperfine induced mixing to be operative at zero magnetic field so preventing clear P-SB, i.e., $E_{ST}<E_N$ where $E_{ST}$ is the energy separation between the relevant singlet and triplet states, and $E_N=g\mu_B\sqrt{<\Delta|B_N|^2>}$~ 0.1μeV is appropriate for ~$10^6$ fluctuating (but unpolarized) nuclear spins [9].

A convenient and well established method to turn-off the hyperfine induced mixing and restore P-SB is to apply a weak magnetic field [9]. The charging diagrams shown in Figs. 3(a) and 3(b) are now re-measured with an in-plane magnetic field B=100mT applied [Figs. 3(c) and 3(d)]. For positive $V_{sd}$, as expected, there is little difference between Fig. 3(c) and Fig. 3(a). In contrast, for negative $V_{sd}$, the current is now strongly suppressed (<1pA) due to P-SB inside the trapezoid region (bounded in red) at the base of the triangle in Fig. 3(d) [contrast with Fig. 3(b) where there is only some current suppression in this region], and this is ultimately because the Zeeman energy splitting ~2.5μeV is much larger than $E_N$ [9,15].

The authors thank K. Ono, S. Sasaki, T. Kodera and A. Shibatomi for fruitful discussions. Part of this work is financially supported by JSPS Grant-in-Aid for Scientific Research S (No. 19104007), MEXT Grant-in-Aid for Scientific Research on Innovative Areas (21102003), Funding Program for World-Leading Innovative R&D on Science and Technology(FIRST), and DARPA QuEST grant HR0011-09-1-0007.

**[References]**

a)Present address: JST, ERATO, Hirayama nuclear spin electronics project, 468-15 Aramaki Aza Aoba, Aoba-ku, Sendai 980-0845, Japan; Electric address: hatano@ncspin.jst.go.jp

1. D. Loss and D. P. DiVincenzo, Phys. Rev. A **57**, 120 (1998).

2. R. Hanson, L. P. Kouwenhoven, J. R. Petta, S. Tarucha, and L. M. K. Vandersypen, Rev. Mod. Phys. **79**, 1217 (2007).

3. J. M. Elzerman, R. Hanson, J. S. Greidanus, L. H. Willems van Beveren, S. De Franceschi, L. M. K. Vandersypen, S. Tarucha, and L. P. Kouwenhoven, Phys. Rev. B **67**, 161308 (2003).

4. D. G. Austing, T. Honda, K. Muraki, Y. Tokura, and S. Tarucha, Physica B **249-251**, 206 (1998).

5. T. Hatano, M. Stopa, and S. Tarucha, Science **309**, 268 (2005).

6. T. Kodera, W. G. van der Wiel, K. Ono, S. Sasaki, T. Fujisawa, and S. Tarucha, Physica E **22**, 518 (2003).

7. K. Ono, D. G. Austing, Y. Tokura, and S. Tarucha, Science **297,** 1313 (2002).

8. K. Ono and S. Tarucha, Phys. Rev. Lett. **92**, 256803 (2004).

9. F. H. L. Koppens, J. A. Folk, J. M. Elzerman, R. Hanson, L. H. Willems van Beveren, I. T. Vink, H. P. Tranitz, W. Wegscheider, L. P. Kouwenhoven, and L. M. K. Vandersypen, Science

**309**, 1346 (2005).

10. F. R. Waugh, M. J. Berry, D. J. Mar, R. M. Westervelt, K. L. Campman, and A. C. Gossard, Phys. Rev. Lett. **75**, 705 (1995).

11. W. G. van der Wiel, S. De Franceschi, J. M. Elzerman, T. Fujisawa, S. Tarucha, and L. P. Kouwenhoven, Rev. Mod. Phys. **75**, 1 (2003).

12. The tunnel coupling energy in the upper-left region of Fig. 2(b) is estimated to be 200~400μeV. For the conditions relevant to Fig. 2(c) and Fig. 3, it is much smaller but hard to determine.

13. A. C. Johnson, J. R. Petta, C. M. Marcus, M. P. Hanson, and A. C. Gossard, Phys. Rev. B **72**, 165308 (2005).

14. The current for transitions involving only the (actual as opposed to effective) charge states (1,0), (2,0), (1,1), and (2,1) is quite low so for this reason we have investigated instead transitions between the related charge states (1,2), (2,2), (1,3), and (2,3) where the current is larger.

15. In the sequence of measurements leading to the data shown in Figs. 3 (b) and (d) we also captured data at 0.2T, 0.3T, and 0.4T, and the clear suppression of current in the trapezoid region bounded in red at 0.1T persisted.

**[Figure Captions]**

Fig. 1 (a) Scanning electron microscope image of a control DQD structure. The direction of the in-plane magnetic field is indicated. (b) Schematic cross section (A-B) of the DQD structure.

Fig. 2 Current through DQD structure as a function of $V_{sL}$ and $V_{sR}$ for $V_{sd}=100\mu V$: (a) $V_c=0.4V$ and $V_b=2.0V$; (b) $V_c=0.35V$ and $V_b=1.8V$; (c) $V_c=0.3V$ and $V_b=2.0V$.

Fig. 3 Current through DQD structure as a function of $V_{sL}$ and $V_{sR}$ for (a) positive $V_{sd}=500\mu V$ and (b) negative $V_{sd}=-500\mu V$ in the absence of an applied magnetic field, and for (c) positive $V_{sd}=500\mu V$ and (d) negative $V_{sd}=-500\mu V$ in the presence of an applied in-plane magnetic field $B=100mT$. In all cases, $V_c=0.28V$ and $V_b=2.5V$. Particularly clear in panels (a) and (c), the ~1pA monotonic decrease in current on moving from the base to the tip of the triangular region is a consequence of the details of inelastic tunneling [9,13].

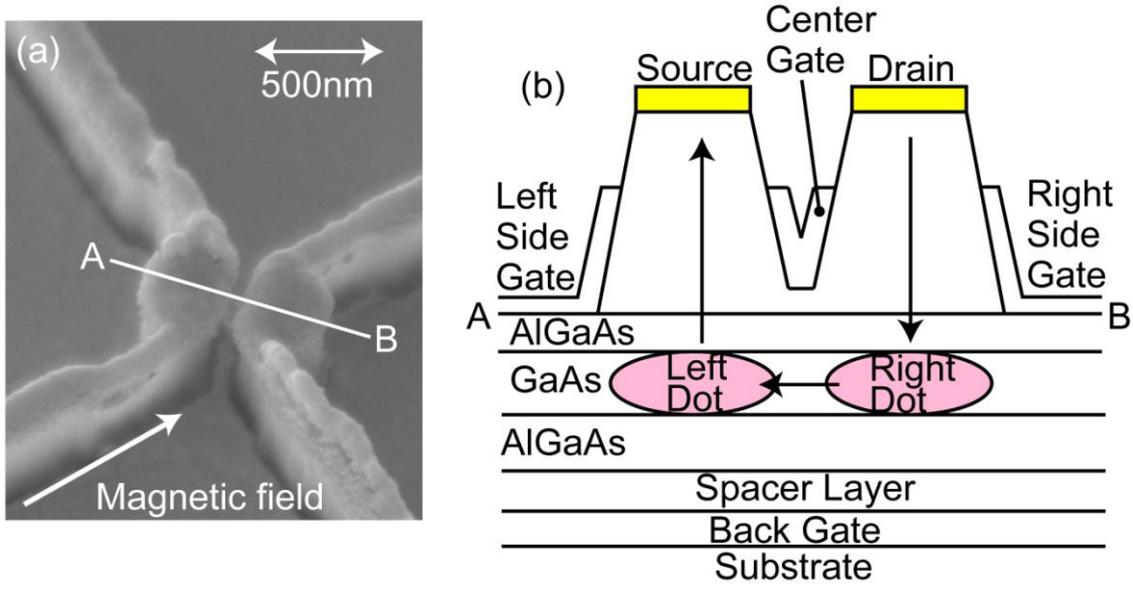

Fig.1

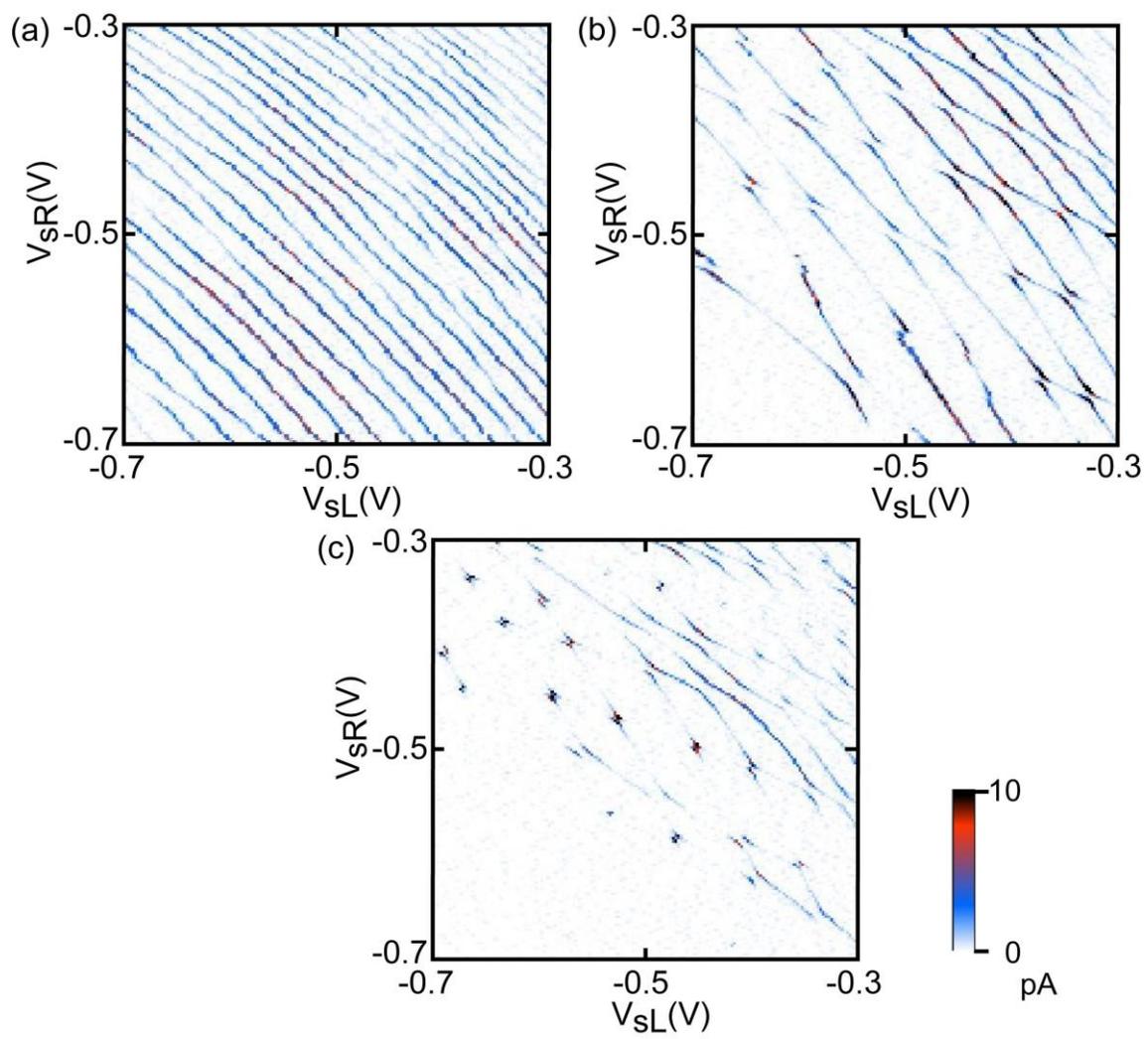

Fig.2

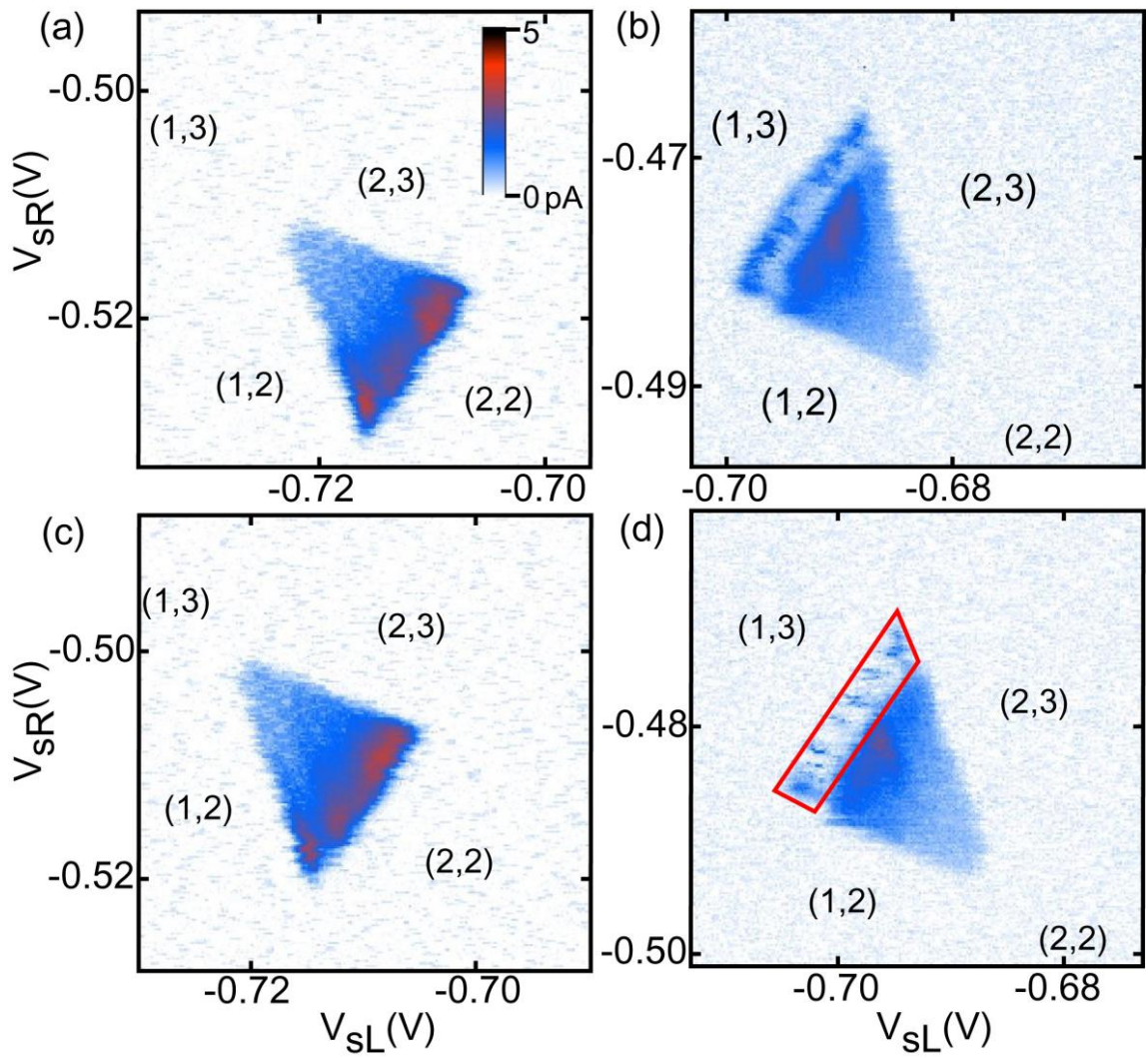

Fig.3